\documentclass[camera,letterpaper,nomarginnotes, 9pt, nonarrowgutter]{jpaper}
\usepackage[table]{xcolor}
\usepackage[T1]{fontenc}
\usepackage{booktabs}
\usepackage[linesnumbered,ruled]{algorithm2e}

\usepackage{listings}
\usepackage{fancyhdr}
\usepackage{datetime}
\usepackage{cite}
\usepackage{amsmath,amssymb,amsfonts}
\usepackage{algorithmic}
\usepackage{graphicx}
\usepackage{textcomp}
\usepackage[table]{xcolor}
\usepackage{tikz}
\usepackage[utf8]{inputenc}
\usepackage[T1]{fontenc}
\usepackage{booktabs} %
\usepackage{setspace}
\usepackage[italic]{mathastext}
\usepackage{array}
\usepackage{titlesec}
\usepackage[normalem]{ulem}
\usepackage{multirow}
\usepackage{multicol}
\usepackage{color}
\usepackage[font={small,bf}]{caption}
\usepackage{float}
\usepackage[font={small,bf}]{subcaption}
\usepackage[linesnumbered,ruled]{algorithm2e}
\usepackage{makecell}
\usepackage{pifont}
\usepackage[]{microtype}

\usepackage[colorinlistoftodos,prependcaption,textsize=small]{todonotes}
\usepackage{marginnote} 

\usepackage{clipboard}
\usepackage{balance}
\usepackage[framemethod=tikz]{mdframed}
\usepackage{xargs}
\usepackage[most]{tcolorbox}

\usepackage[bookmarks=true,breaklinks=true,colorlinks,linkcolor=black,citecolor=blue,urlcolor=black]{hyperref}

\usepackage[resetlabels]{multibib}

\definecolor{denim}{rgb}{0.08, 0.38, 0.74}
\definecolor{darkolivegreen}{rgb}{0.33, 0.42, 0.18}
\definecolor{dgreen}{rgb}{0.00, 0.75, 0.00}
\definecolor{darkpink}{rgb}{0.88, 0.28, 0.54}
\definecolor{forestgreen}{rgb}{0.0, 0.27, 0.13}
\definecolor{amber}{rgb}{1.0, 0.49, 0.0}
\definecolor{lightyellow}{rgb}{0.980, 0.956, 0.623}
\definecolor{lightblue}{rgb}{0.980, 0.956, 0.623}
\definecolor{darkamber}{rgb}{0.5, 0.19, 0.0}
\definecolor{dkgreen}{rgb}{0,0.6,0}
\definecolor{gray}{rgb}{0.5,0.5,0.5}
\definecolor{mauve}{rgb}{0.58,0,0.82}
\definecolor{lightmauve}{rgb}{0.68,0.4,0.92}
\definecolor{chocolate}{rgb}{0.48, 0.25, 0.0}
\definecolor{dollarbill}{rgb}{0.52,0.73,0.4}
\definecolor{dkdkgreen}{rgb}{0,0.45,0}
\definecolor{gfored}{rgb}{0.580, 0.050, 0.211}
\definecolor{darkwarmgray}{rgb}{0.15, 0.050, 0.05}
\definecolor{ups-truck}{rgb}{0.53, 0.28, 0.21}

\definecolor{bestresult}{HTML}{b3e2cd} %
\definecolor{worseresult}{HTML}{f4c7c3} %

\newcommand\rev[1]{{\color{black}{#1}}}
\newcommand\revb[1]{{\color{black}{#1}}}
\newcommand\revc[1]{{\color{black}{#1}}}

\makeatletter
\g@addto@macro{\normalsize}{%
  \setlength{\abovedisplayskip}{2pt plus 1pt minus 1pt}
  \setlength{\belowdisplayskip}{2pt plus 1pt minus 1pt}
  \setlength{\intextsep}{2pt plus 1pt minus 1pt}
  \setlength{\textfloatsep}{3pt plus 1pt minus 1pt}
  \setlength{\dbltextfloatsep}{3pt plus 1pt minus 1pt}
  \setlength{\skip\footins}{4pt plus 1pt minus 1pt}
}
\def\BibTeX{{\rm B\kern-.05em{\sc i\kern-.025em b}\kern-.08em
    T\kern-.1667em\lower.7ex\hbox{E}\kern-.125emX}}

\def\UrlBreaks{\do\/\do-\/\do.\/\do:}

\expandafter\def\expandafter\UrlBreaks\expandafter{\UrlBreaks
  \do\a\do\b\do\c\do\d\do\e\do\f\do\g\do\h\do\i\do\j
  \do\k\do\l\do\m\do\n\do\o\do\p\do\q\do\r\do\s\do\t
  \do\u\do\v\do\w\do\x\do\y\do\z\do\A\do\B\do\C\do\D
  \do\E\do\F\do\G\do\H\do\I\do\J\do\K\do\L\do\M\do\N
  \do\O\do\P\do\Q\do\R\do\S\do\T\do\U\do\V\do\W\do\X
  \do\Y\do\Z}
  
\newcommand{\squishlist}{
 \begin{list}{$\circ$}
  { \setlength{\itemsep}{0pt}
     \setlength{\parsep}{0pt}
     \setlength{\topsep}{0pt}
     \setlength{\partopsep}{0pt}
     \setlength{\leftmargin}{1em}
     \setlength{\labelwidth}{1em}
     \setlength{\labelsep}{0.5em} } }

\newcommand{\squishsublist}{
\begin{list}{$\rightarrow$}
 { \setlength{\itemsep}{0pt}
    \setlength{\parsep}{0pt}
    \setlength{\topsep}{-10em}
    \setlength{\partopsep}{-3pt}
    \setlength{\leftmargin}{1em}
    \setlength{\labelwidth}{1em}
    \setlength{\labelsep}{0.5em} } }

\newcommand{\squishend}{
  \end{list}  }

\newcommand{\head}[1]{\noindent\textbf{#1.}} %

\newcommand{\circled}[1]{{\tikz[baseline=(char.base)]{\node[shape=circle,inner sep=1.3pt,fill=black, text=white] (char) {\small \textbf{#1}};}}}

\usepackage{titlesec}
\titlespacing*{\section}{0pt}{0.3ex}{0.1ex} %
\titlespacing*{\subsection}{0pt}{0.3ex}{0.3ex} %
\titlespacing*{\subsubsection}{0pt}{0.3ex}{0.3ex} %

\newcommand\rs{Rawsamble\xspace}
\newcommand\rsltitle{Rawsamble: Overlapping Raw Nanopore Signals\\using a Hash-based Seeding Mechanism\xspace}

\newcommand{\rsrelease}{\href{https://github.com/CMU-SAFARI/RawHash}{https://github.com/CMU-SAFARI/RawHash}\xspace}

\newcommand{\rht}{{RawHash2}\xspace}

\newcommand\rsavgthr{$1{,}533{,}035$\xspace}

\newcommand\rsavgcpufcpu{$5.46\times$\xspace}

\newcommand\rsavgelfcpu{$5.01\times$\xspace}
\newcommand\rsmaxelfcpu{$23.10\times$\xspace}
\newcommand\rsavgpeakfcpu{$5.74\times$\xspace}
\newcommand\rsmaxpeakfcpu{$22.00\times$\xspace}
\newcommand\rsavgcpuhcpu{$18.21\times$\xspace}

\newcommand\rsavgelhcpu{$30.36\times$\xspace}

\newcommand\rsavgpeakhcpu{$6.67\times$\xspace}

\newcommand\rsavgelhgpu{$0.51\times$\xspace}

\newcommand\rsavgpeakhgpu{$1.49\times$\xspace}

\newcommand\rsavgshared{$27.09\%$\xspace}
\newcommand\rsavgru{$12.87\%$\xspace} %
\newcommand\rsavgmmu{$60.04\%$\xspace} %

\newcommand{\citenanopore}{menestrina_ionic_1986,cherf_automated_2012,manrao_reading_2012,laszlo_decoding_2014,deamer_three_2016,kasianowicz_characterization_1996,meller_rapid_2000,stoddart_single-nucleotide_2009,laszlo_detection_2013,schreiber_error_2013,butler_single-molecule_2008,derrington_nanopore_2010,song_structure_1996,walker_pore-forming_1994,wescoe_nanopores_2014,lieberman_processive_2010,bezrukov_dynamics_1996,akeson_microsecond_1999,stoddart_nucleobase_2010,ashkenasy_recognizing_2005,stoddart_multiple_2010,bezrukov_current_1993,zhang_single-molecule_2024}

\newcommand{\citebasecallnanodnn}{cavlak_targetcall_2024,xu_fast-bonito_2021,peresini_nanopore_2021,boza_deepnano_2017,boza_deepnano-blitz_2020,oxford_nanopore_technologies_dorado_2024,oxford_nanopore_technologies_guppy_2017,lv_end--end_2020,singh_rubicon_2024,zhang_nanopore_2020,xu_lokatt_2023,zeng_causalcall_2020,teng_chiron_2018,konishi_halcyon_2021,yeh_msrcall_2022,noordijk_baseless_2023,huang_sacall_2022,miculinic_mincall_2019}
\newcommand{\citebasecallnanohmm}{loman_complete_2015,david_nanocall_2017,timp_dna_2012,schreiber_analysis_2015}

\newcommand{\citesignalanalysis}{bao_squigglenet_2021,loose_real-time_2016,zhang_real-time_2021,kovaka_targeted_2021,senanayake_deepselectnet_2023,sam_kovaka_uncalled4_2024,lindegger_rawalign_2023,firtina_rawhash_2023,firtina_rawhash2_2024,shih_efficient_2023,sadasivan_rapid_2023,dunn_squigglefilter_2021,shivakumar_sigmoni_2024,sadasivan_accelerated_2024,gamaarachchi_gpu_2020,samarasinghe_energy_2021}
\newcommand{\citesignalanalysismapped}{loose_real-time_2016,zhang_real-time_2021,kovaka_targeted_2021,lindegger_rawalign_2023,firtina_rawhash_2023,firtina_rawhash2_2024,shih_efficient_2023,dunn_squigglefilter_2021,shivakumar_sigmoni_2024,sadasivan_accelerated_2024,gamaarachchi_gpu_2020,samarasinghe_energy_2021}

\newcommand{\versionnum}[0]{1.1}
\fancypagestyle{firststyle}{
    \fancyhead[C]{\textcolor{blue}{\emph{Version \versionnum~---~\today, \xxivtime \ UTC}}}%
    \fancyfoot[C]{\thepage}
}

\def\BibTeX{{\rm B\kern-.05em{\sc i\kern-.025em b}\kern-.08em
    T\kern-.1667em\lower.7ex\hbox{E}\kern-.125emX}}
    
\newcites{supp}{Supplementary References}

\begin{document}

\title{\rsltitle}

\newcommand{\affila}[0]{\small {$^1$}}
\newcommand{\affilb}[0]{\small {$^2$}}
\newcommand{\affilc}[0]{\small {$^3$}}
\newcommand{\affild}[0]{\small {$^4$}}
\newcommand{\affile}[0]{\small {$^5$}}
\author{
\vspace{-18pt}\\%
{Can Firtina\affila{}$^,$\affilb{}}\quad%
{Maximilian Mordig\affila{}$^,$\affilc{}}\quad%
{Harun Mustafa\affila{}$^,$\affild{}$^,$\affile{}}\quad%
{Sayan Goswami\affila{}}\quad%
\vspace{-1pt}\\%
{Nika Mansouri Ghiasi\affila{}}\quad%
{Stefano Mercogliano\affila{}}\quad%
{Furkan Eris\affila{}}\quad%
{Joël Lindegger\affila{}}\quad%
\vspace{-1pt}\\%
{Andre Kahles\affila{}$^,$\affild{}$^,$\affile{}}\quad%
{Onur Mutlu\affila{}}\quad%
\vspace{-1pt}\\%
\affila\emph{ETH Zurich}%
\qquad\quad%
\affilb\emph{University of Maryland, College Park}%
\qquad\quad%
\affilc\emph{Max Planck Institute for Intelligent Systems}%
\vspace{-1pt}\\%
\affild\emph{University Hospital Zurich}%
\qquad\quad%
\affile\emph{Swiss Institute of Bioinformatics}%
\vspace{-12pt}
}

\maketitle
\pagestyle{plain}
\thispagestyle{plain}
\setstretch{0.98}
\begin{abstract}
{\noindent \textbf{Abstract:} Raw nanopore signal analysis is a common approach in genomics to provide fast and resource-efficient analysis without translating the signals to bases (i.e., without basecalling). However, existing solutions cannot interpret raw signals directly if a reference genome is unknown due to a lack of accurate mechanisms to handle increased noise in pairwise raw signal comparison. Our goal is to enable the direct analysis of raw signals without a reference genome. To this end, we propose \rs, the \textit{first} mechanism that can identify regions of similarity between all raw signal pairs, known as \textit{all-vs-all overlapping}, using a hash-based search mechanism.

We use these overlaps to construct \textit{de novo} assembly graphs with an existing assembler, miniasm, off-the-shelf. To our knowledge, these are the first \textit{de novo} assemblies ever constructed directly from raw signals without basecalling. Our extensive evaluations across multiple genomes of varying sizes show that \rs provides a significant speedup (on average by \rsavgelfcpu and up to \rsmaxelfcpu) and reduces peak memory usage (on average by \rsavgpeakfcpu and up to by \rsmaxpeakfcpu) compared to a conventional genome assembly pipeline using the state-of-the-art tools for basecalling (Dorado's fastest mode) and overlapping (minimap2) on a CPU. We find that around one-third of \rs’s overlapping pairs are also found by minimap2. We find that when we use overlapping reads from \rs, we can construct unitigs that are 1) as accurate as those built from minimap2’s overlaps and 2) up to half a chromosome in length (e.g., 2.3 million bases for \textit{E. coli}).
\\\textbf{Availability and Implementation:} \rs is available at \rsrelease. We also provide the scripts to fully reproduce our results on our GitHub page. \\
}
\end{abstract}

\section{Introduction} \label{rs:sec:introduction}
Nanopore sequencing technology~\cite{\citenanopore} can sequence long nucleic acid molecules\rev{, called \emph{reads},} of up to a few million bases at high throughput~\cite{jain_nanopore_2018,pugh_current_2023,senol_cali_nanopore_2019}. As a molecule moves through a tiny \emph{nanopore}, ionic current measurements, called \emph{raw signals}, are generated~\cite{deamer_three_2016}.

\revc{Nanopore sequencing provides two unique key benefits.}
First, nanopore sequencing enables stopping the sequencing of single reads or the entire sequencing run early, known as \emph{adaptive sampling} or \emph{selective sequencing}~\cite{loose_real-time_2016},
while raw signals are generated and analyzed during sequencing, called \emph{real-time analysis}. Adaptive sampling can substantially reduce the sequencing time and cost by avoiding unnecessary sequencing.
Second, compact nanopore sequencing devices enable on-site portable sequencing and analysis, which can be coupled with real-time analysis~\cite{bloemen_development_2023}.

Existing works that analyze raw nanopore signals~\cite{\citebasecallnanodnn,\citebasecallnanohmm,\citesignalanalysis, eris_rawbench_2025} mainly utilize \rev{complex} deep learning mechanisms~\cite{\citebasecallnanodnn} to translate these signals into nucleotides, a process called \emph{basecalling}.
Basecalling mechanisms usually 1)~are designed to use large chunks of raw signal data for accurate analysis~\cite{zhang_real-time_2021}, 2)~have high computational requirements~\cite{shih_efficient_2023,senol_cali_nanopore_2019}, \revb{and 3)~oblivious to the similarity information between raw signals that lead to redundant basecalling on these similar regions. Existing basecalling techniques and the lack of understanding of similarity between raw signals can impose limitations to enable 1)~accurate real-time analysis~\cite{zhang_real-time_2021} and 2)~portable sequencing with constrained resources~\cite{shih_efficient_2023} and prevent future work on designing a new class of basecalling techniques that can utilize the similarity information between raw signals.}

To fully utilize the unique benefits of nanopore sequencing, it is necessary to analyze raw signals with 1)~high accuracy and low latency for adaptive sampling and 2)~low resource usage for portability and efficiency. To achieve this, several mechanisms focus on analyzing raw nanopore signals \emph{without} basecalling~\cite{\citesignalanalysis} by mapping these signals to a reference genome.\footnote{We use the \emph{raw signal analysis} term specifically for these mechanisms in the remainder of the paper.}
\revb{Although raw nanopore signal mapping is widely studied~\cite{\citesignalanalysismapped}, \emph{none} of these works \emph{without} a reference genome to identify similarities directly between reads, called \emph{all-vs-all overlapping}~\cite{li_minimap_2016, senol_cali_nanopore_2019, firtina_blend_2023}.}

Although identifying all-vs-all overlapping between raw nanopore signals can be promising to enable \revb{\emph{new} directions for \emph{both} basecalling mechanisms (as we discuss in Section~\ref{rs:sec:discussion}) and raw signal analysis (e.g., \emph{de novo} assembly construction as we show in Section~\ref{rs:subsec:assembly})}, it is challenging to do so for several reasons~\cite{deamer_three_2016,bhattacharya_molecular_2012,kawano_controlling_2009,smeets_noise_2008}.
\rev{First, similarities between a pair of signals must be identified accurately when \emph{both} raw signals are noisy} as compared to identifying similarities between a noisy raw signal and an accurate signal generated from a reference genome~\cite{zhang_real-time_2021, firtina_rawhash_2023, firtina_rawhash2_2024, lindegger_rawalign_2023}. Converting reference genomes to their expected raw signal values is free from certain types of noise that raw nanopore signals contain (e.g., stochastic signal fluctuations~\cite{deamer_three_2016}, variable speed of DNA molecules moving through nanopores~\cite{bhattacharya_molecular_2012,kawano_controlling_2009}, \revb{and raw reads containing multiple split molecules)}, while raw signals are \emph{not} free from such noise.

Second, existing raw signal analysis works lack the mapping strategies typically used in all-vs-all overlapping, such as reporting multiple mappings (i.e., overlaps) of a read to many reads. This is because these works are mainly designed to stop the mapping process as soon as there is an accurate mapping for a read to minimize the unnecessary sequencing~\cite{kovaka_targeted_2021}. For all-vs-all overlapping, pairwise mappings of a read to many reads (instead of a single mapping) must be reported while avoiding certain trivial cyclic pairwise mappings to construct an assembly~\cite{li_minimap_2016}.

Third, read overlapping can increase the space requirements for storing and using indexing. This is because a read set is likely to be sequenced such that its overall number of bases is larger than the number of bases in its corresponding genome. Such an increased index size can raise the computational and space demands for read overlapping. It is essential to provide high-throughput and scalable computation to enable real-time downstream analysis for future work (as discussed in Section~\ref{rs:sec:discussion}).

\textbf{Our goal} is to enable 1)~raw signal analysis \emph{without} a reference genome and 2)~new use cases with raw nanopore signal analysis by addressing the challenges of accurate and fast all-vs-all overlapping of raw nanopore signals. To this end, we propose \emph{\rs}, the first mechanism that enables fast and accurate overlap finding between raw nanopore signals. The key idea in \rs is to re-design the existing state-of-the-art hash-based seeding mechanism~\cite{firtina_rawhash_2023, firtina_rawhash2_2024} for raw signals with more effective noise reduction techniques and useful outputting strategies to find all overlapping pairs accurately, which we explain in three key steps.

First, to enable identifying similarities between a pair of noisy raw signals accurately, \rs filters raw signals to select those substantially distinct from their surrounding signals. Such non-distinct and adjacent signals are usually the result of a certain error type in the analysis, known as \emph{stay errors}~\cite{shivakumar_sigmoni_2024, firtina_rawhash_2023,zhang_real-time_2021}.
Although similar filtering strategies~\cite{shivakumar_sigmoni_2024, firtina_rawhash_2023,zhang_real-time_2021} are exploited when mapping raw signals to reference genomes, \rs performs a more aggressive filtering to avoid storing the erroneous portions of signals in the index to enable accurate similarity identification from the sufficiently distinct regions of signals.
Second, to find multiple overlaps for a read, \rs identifies highly accurate chains from seed matches based on their chaining scores and reports \emph{all} of these chains as mappings, as opposed to choosing solely the best mapping as determined by weighted decisions among all such chains~\cite{firtina_rawhash2_2024}.
Third, to prevent trivial cycles between a pair of overlapping reads, \rs ensures that only one of the overlapping reads in each pair is always chosen as a query sequence, and the other is always chosen as a target sequence based on a deterministic ordering mechanism between these reads. These steps enable \rs to find overlaps between raw signals accurately and quickly.

\rs makes the following key contributions:

\begin{itemize}
    \item We propose the \emph{first} mechanism that can find all-vs-all overlapping of raw signals without basecalling to enable new use cases for raw signal analysis, such as \emph{assembly from overlapping} raw signals \rev{and improving existing basecalling mechanisms}.
    \item We show that identifying overlaps with \rs 1)~is faster (on average \rsavgelfcpu and up to \rsmaxelfcpu) and 2)~reduces peak memory requirements (on average \rsavgpeakfcpu and up to \rsmaxpeakfcpu) compared to the computational pipeline that includes the state-of-the-art basecalling running on a CPU with its fastest model followed by minimap2 to find overlaps. We evaluate the trade-offs between speed, accuracy, and hardware resources by running the basecaller on a CPU and GPU with fast and high-accuracy models.
    \item We show that \rsavgshared of overlapping pairs that \rs generates are identical to the overlapping pairs generated by minimap2.
    \item We report the first \emph{de novo} \rev{assembly graphs} ever constructed directly from raw signal overlaps without basecalling. We show that we can construct long unitigs up to 2.3 million bases in length for \textit{E. coli}, which constitutes half the length of the genome. \revb{We find that these assemblies constructed from raw signal overlaps are either as accurate or more accurate than those created by minimap2 overlaps.}
    \item \rev{We identify new use cases that can be enabled by using overlapping raw signals and constructing assemblies from them \revb{and discuss the advantages of generating them,} such as for improving the accuracy and performance of basecalling. We discuss current limitations and challenges to enable these new use cases as future work.}
\end{itemize}

\section{Methods} \label{rs:sec:methods}
\subsection{Overview} \label{rs:subsec:met_overview}
\begin{figure}[tbh]
  \centering
  \includegraphics[width=\columnwidth]{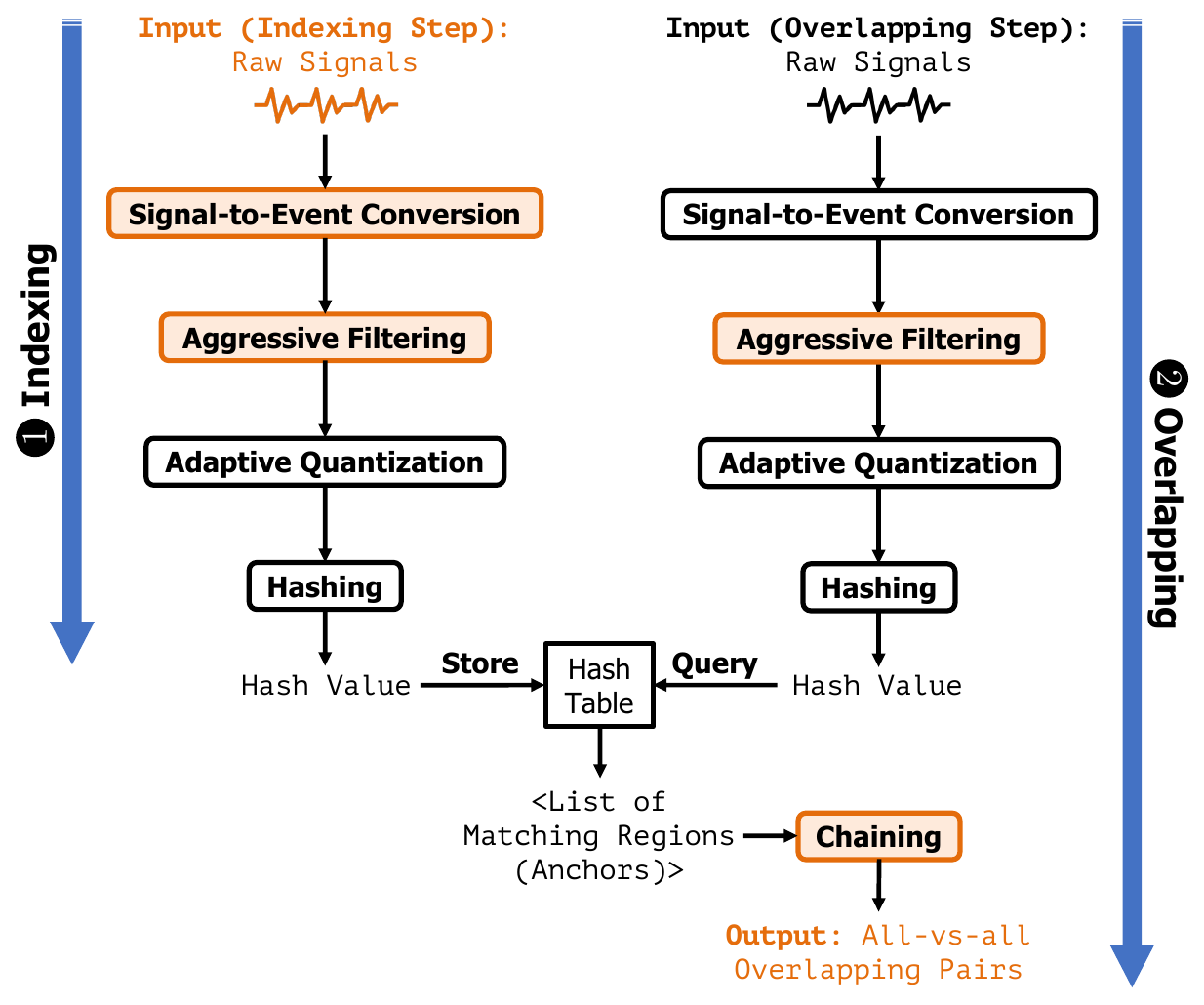}
  \caption{Overview of \rs. We use colors for the inputs, steps, and outputs to highlight the parts that \rs modifies over \rht.}
  \label{rs:fig:overview}
\end{figure}

\rs is a mechanism to find overlapping pairs between raw nanopore signals (i.e., \emph{all-vs-all overlapping}), \rev{which can be used by existing assemblers} to construct \emph{de novo} assembly graphs without basecalling, as shown in Figure~\ref{rs:fig:overview}. To achieve this, \rs builds on the state-of-the-art raw signal mapper, \rht~\cite{firtina_rawhash_2023, firtina_rawhash2_2024}. We provide the overview of \rht in Supplementary Section~\ref{rs:suppsec:rht_overview}, \revb{where we explain the details related to processing raw signals such as generating \emph{events} after the \emph{segmentation} and generating hash values from these segmented raw signals (i.e., events). In the remainder of the paper, we use \emph{raw signals} to refer to these events.}

\rs extends \rht to support all-vs-all raw signal overlapping in four key steps.
First, to enable efficient and accurate indexing from the noisy raw signals (\circled{1}), \rs aggressively filters the raw input after the signal-to-event conversion to avoid nanopore-related errors.
Second, to improve the accuracy of overlapping (\circled{2}), \rs adjusts the minimum chaining score to avoid false chains, ensuring that only high-confidence overlaps are considered.
Third, to enable finding useful and long connections from many overlaps, \rs adjusts the output strategy such that 1)~\emph{all} chains rather than only the best chain are reported and 2)~cyclic overlaps are avoided.
\rev{Fourth, \rs enables the use of existing \emph{de novo} assemblers \rev{off-the-shelf}, such as miniasm~\cite{li_minimap_2016}, by providing the overlap information in a standardized format these assemblers use.} 

\subsection{Constructing an Index from Noisy Raw Signals via Aggressive Filtering} \label{rs:subsec:met_indexing}
\begin{figure}[tbh]
  \centering
  \includegraphics[width=\columnwidth]{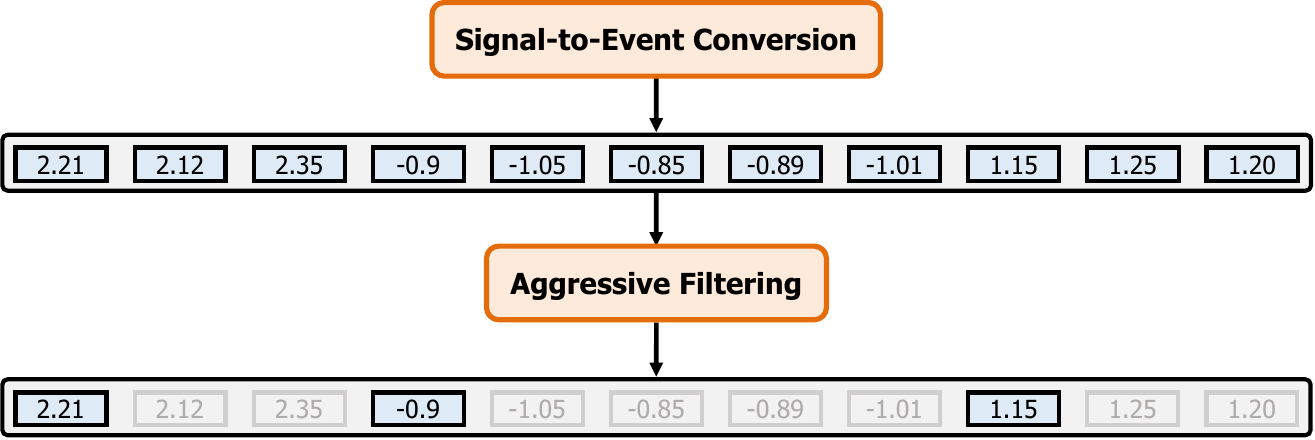}
  \caption{Filtering in \rs. Values in gray boxes show the filtered signals as their values are close to the previous signal (in a blue box) that is not filtered out.}
  \label{rs:fig:filter}
\end{figure}

\rs identifies overlapping regions between raw nanopore signals using a hash-based seeding mechanism that operates in two steps.
First, to enable quick matching between raw signals, \rs enables constructing an index directly from raw nanopore signals instead of using an existing reference genome sequence. While converting reference genomes to their expected raw signal values is mainly free from certain types of noise that raw nanopore signals contain (e.g., stochastic signal fluctuations~\cite{deamer_three_2016} and variable speed of DNA molecules moving through nanopores~\cite{bhattacharya_molecular_2012,kawano_controlling_2009}), noise in raw nanopore signals can cause challenges to find accurate matches between raw signals when these signals are stored in an index.
Second, to reduce noise stored in the hash tables and enable accurate similarity identification when both signals are noisy, \rs aggressively filters raw signals as shown in Figure~\ref{rs:fig:filter}. The filtering mechanism iteratively compares two adjacent signals, $s_i$ and $s_j$, and removes the second signal, $s_j$, if the absolute difference between the two adjacent signals is below a certain threshold $T$. This filtering generates a list of filtered signals that aggressively aims to reduce the impact of stay errors during sequencing. Although similar filtering approaches are used in prior works~\cite{zhang_real-time_2021, firtina_rawhash_2023, firtina_rawhash2_2024} to reduce the stay errors (e.g., by aiming to perform homopolymer compression of raw signals~\cite{shivakumar_sigmoni_2024}), \rs employs a substantially larger threshold (i.e., aggressive) for filtering to minimize noise both in the index and during mapping.
By applying the aggressive filtering technique, \rs ensures that only high-quality, informative events are used in indexing to improve the accuracy and efficiency of the overall overlapping mechanism when both signals are noisy.

\subsection{Adjusting the Chaining Mechanism for Overlapping}
To reduce the number of chains that do not result in mapping (i.e., false chains) and construct longer chains, \rs adjusts the chaining mechanism~\cite{li_minimap2_2018, firtina_rawhash2_2024} in two ways.
First, \rs constructs chains between seed matches (i.e., anchors) with longer gaps by adjusting the maximum gap length between anchors. This adjustment is needed because the filtering mechanism results in a sparser list with potentially long gaps between signals.
Second, to ensure that only high-confidence chains are considered as overlaps among many pairwise overlapping regions, \rs sets a higher minimum chaining score for overlapping than mapping, which effectively filters out spurious matches. 
These adjustments to the chaining mechanism enable \rs to construct long and accurate chains between noisy and sparse raw signals, improving the overall sensitivity of the overlapping process and further downstream analysis such as \emph{de novo} assembly construction.

\subsection{Adjusting the Mapping Strategy}
\rs adjusts the mapping strategy used in \rht in two ways.
First, \rs generates pairwise mappings of a read to many reads (instead of a single mapping per read). To achieve this, \rs identifies all valid chains based on their chaining scores and reports all such chains between raw signals. This enables the overlapping of a single raw signal with multiple other raw signals.
Second, \rs filters out \emph{trivial} cyclic overlaps between two reads. To do so, \rs avoids reporting overlapping signal pairs both as \emph{query} and \emph{target} in the mapping output, which can complicate the \emph{de novo} assembly construction process~\cite{li_minimap_2016, simpson_abyss_2009, i_sovic_approaches_2013}. To avoid these trivial cyclic overlaps, \rs implements a pre-defined and deterministic ordering between raw signals (e.g., based on the lexicographic ordering of read names). By comparing raw signals deterministically, \rs guarantees that only one of each pair of overlapping reads is processed as a query sequence while the other is treated as a target sequence.
Reporting all chains while avoiding cyclic overlaps between raw signals allows for a comprehensive representation of the overlapping regions, which is useful for constructing accurate and long assemblies.

\head{Output Format}
\rs provides the overlapping information between raw signals using the Pairwise mApping Format (PAF)~\cite{li_minimap_2016}.
To construct an assembly graph from the overlapping information that \rs generates, any assembler that takes PAF files as input can be used, including miniasm~\cite{li_minimap_2016}.

\section{Results} \label{rs:sec:results}
\subsection{Evaluation Methodology} \label{rs:subsec:evaluation}
We implement the improvements we propose in \rs on the \rht implementation~\cite{firtina_rawhash2_2024}. Similar to \rht and minimap2~\cite{li_minimap2_2018}, \rs provides the mapping information in the standard Pairwise mApping Format (PAF)~\cite{li_minimap_2016}.
We basecall the signals on two different hardware setups using Dorado's 1)~high accuracy (HAC) model with an NVIDIA RTX A6000 GPU~\cite{nvidia_a6000_2024}, and 2)~fast (Fast) and HAC models with an Intel Xeon Gold 6226R CPU~\cite{intel_xeon6226R_2024}. \revb{We include Dorado's results on a CPU to provide a fair comparison on the same hardware platform, as providing the GPU implementation for Rawsamble is a future work.} We use POD5 files for all datasets, as suggested by Dorado~\cite{oxford_nanopore_technologies_dorado_2024} for optimal performance.
Since no prior works can overlap reads using raw signals, we compare \rs to minimap2, the current state-of-the-art read overlapper for basecalled sequences using the datasets shown in Table~\ref{rs:tab:dataset}. \revb{Although the R9.4.1 nanopore chemistry is deprecated, these raw datasets are commonly stored and basecalled for downstream analysis. We include datasets from both chemistries to show the feasibility and limitations of \rs for each chemistry.} \revb{For the R10.4.1 dataset, we use only the raw signals that contain a single molecule (i.e., non-chimeric multiple molecules can be contained in a single nanopore raw signal and later split by basecallers into multiple reads)}.

\begin{table}[htb]
\centering
\caption{Details of datasets used in our evaluation.}
\resizebox{\linewidth}{!}{

\begin{tabular}{@{}clllrrrr@{}}\toprule
& \textbf{Organism}     & \textbf{Chemistry}  &\textbf{Reads} & \textbf{Bases}  & \textbf{Avg. Read} & \textbf{Estimated}           & \textbf{SRA}       \\
& \textbf{}             & \textbf{Model}    &\textbf{(\#)}  & \textbf{(\#)}   & \textbf{Length}   & \textbf{Coverage ($\times$)} & \textbf{or DOI} \\\midrule
D1 & \emph{E. coli}     & R9.4         & 353,956       & 2,360M          & 6,668             & 451$\times$                  & ERR9127551         \\\midrule
D2 & \emph{Yeast}       & R9.4          & 49,992        & 380M            & 7,609             & 31$\times$                   & SRR8648503         \\\midrule
D3 & \emph{Green Algae} & R9.4      & 30,012        & 622M            & 20,728 & 5.6$\times$                  & ERR3237140         \\\midrule
D4 & \emph{Human}       & R9.4          & 270,012       & 1,776M & 6,579          & 0.6$\times$                  & FAB42260           \\\midrule
D5 & \emph{E. coli} & R10.4.1      & 6,412,132        & 9,400M            & 1,466            & 1,797$\times$                  & 10.26188/25521892         \\\bottomrule
\multicolumn{8}{l}{Base counts in millions (M). Coverage is estimated using corresponding reference genomes.}\\
\multicolumn{8}{l}{All of the datasets are basecalled using Dorado (HAC).}\\
\end{tabular}
}
\label{rs:tab:dataset}
\end{table}

We evaluate computational requirements in terms of overall run time with 64 CPU threads, peak memory usage, and \revc{the estimated cloud computing costs to execute each tool}. We use 64 CPU threads since the average thread utilization of CPU-based basecalling is around 64. We report the elapsed time, CPU time (when using only CPUs without GPUs), and peak memory usage of 1)~\rs and 2)~basecalling followed by minimap2 (\texttt{Dorado + Minimap2}). For \rs, we additionally report throughput (average number of signals analyzed per second per single CPU thread) to provide insights about its capabilities for real-time analysis. To measure performance and peak memory usage, we use the \texttt{time -v} command in Linux when running \rs, minimap2, and Dorado. For average speedup and memory comparisons of \rs against other methods, we use the geometric mean to reduce the impact of outlier data points on the average calculation.

We evaluate \rs based on two use cases: 1)~read overlapping and 2)~\emph{de novo} assembly \rev{by constructing assembly graphs}. We generate read overlaps from 1)~raw signals using \rs and 2)~basecalled sequences (with Dorado's HAC model) of corresponding strands of raw signals using minimap2. To evaluate \rev{the accuracy} of all-vs-all overlapping, we calculate the ratio of overlapping pairs 1)~shared by both \rs and minimap2, 2)~unique to \rs, and 3)~unique to minimap2.
For \emph{de novo} assembly, we use off-the-shelf miniasm~\cite{li_minimap_2016} to generate assembly graphs from the read overlaps that \rs and minimap2 generate. We use miniasm because it enables us to evaluate the impact of overlaps that \rs and minimap2 find by \emph{using the same assembler} for both of them.
To identify the roofline in terms of the assembly contiguity given a dataset, we use a highly accurate assembler as \emph{gold standard} for our evaluations. To do this, we use Dorado's most accurate model (i.e., SUP) to construct highly accurate reads and use Flye~\cite{kolmogorov_assembly_2019} to construct assemblies from these reads, as suggested in the guidelines by ONT~\cite{oxford_nanopore_technologies_guidelines_2022}. This approach enables us to evaluate the gap between the assemblies that \rs generates and the golden standard assembly that can be constructed from the same datasets.

To evaluate the contiguity of assemblies constructed from the overlaps that \rs and minimap2 generate, we use several metrics. These metrics are 1)~the total length of unitigs (Total Length), 2)~the number of nucleotides in the largest assembly graph component (Largest Comp.), 3)~unitig length at which 50\% of the assembly is covered by unitigs of equal or greater length (N50), 4)~area under the Nx curve to generate a more robust evaluation of contiguity (auN)~\cite{li_aun_2020}, 5)~longest unitig length, and 6)~the number of unitigs. \revb{We \emph{estimate} the basecalled read length of raw signals based on the sample frequency (i.e., the number of signals produced every second) and the translocation speed (i.e., the number of bases estimated to pass through a nanopore in a single second). Estimating the read length based on these parameters is a common and relatively consistent approach widely adopted in previous related works~\cite{kovaka_targeted_2021,zhang_real-time_2021, firtina_rawhash_2023, firtina_rawhash2_2024}.} We use Bandage~\cite{wick_bandage_2015} to visualize these assemblies. We note that our contiguity evaluations for \rs and minimap2 are based on the assembly graphs constructed by miniasm.

\revb{To evaluate the accuracy of the constructed graphs from \rs and minimap2 overlaps, we perform read mapping from their corresponding basecalled reads to their corresponding reference genomes. For each unitig generated in a graph, we measure how well its reads line up on the reference genome in four steps.
First, we list the reads of the unitig in the exact order given by the assembly graph.
Second, for each read, we collect every alignment of that read to the reference genome and create an anchor $\langle i,s\rangle$, where $i$ is the ordinal position of the read in the unitig and $s$ is the reference start coordinate.
Third, we find the longest strictly increasing sequence of anchors that satisfies two constraints: 1)~all anchors in the chain must map to the same reference chromosome and in the same left-to-right order, and 2)~the start position of anchor~$j{+}1$ must lie in the interval $[s_j,\;s_j+\mathrm{readlen}_j+\delta)$, where $\delta$\,bp is a tolerance threshold for the range. Our dynamic programming implementation on this chaining algorithm provides the number of reads in the best chain, $k$, where the goal of the best chain is to maximize $k$. Fourth, we record the ratio $R = k/n$ where $n$ is the total number of reads in the unitig (unitigs containing no more than two reads are skipped). Summing $k$ and $n$ over all multi-read unitigs gives overall totals $\text{all\_k}$ and $\text{all\_n}$. We estimate the accuracy of the assembly graph as $100\times\text{all\_k}/\text{all\_n}$, i.e.\ the percentage of all reads within the same unitig that can be placed in a consistent, chromosome-specific, nearly contiguous order with respect to the reference genome, which we report as \texttt{Chained Read Percentage}. The goal of generating these percentages is to provide our best attempt for a \emph{meaningful comparison} between assembly graphs in terms of their \emph{estimated} accuracy. \textbf{We cannot align the assembled raw signals in an assembly graph to a reference genome to measure their actual identity, since assembled raw signals cannot be reported.} We leave 1)~constructing the assembled signal sequence (i.e., spelling the assembly from the graph), 2)~designing a consensus mechanism for error correction, and 3)~evaluating the accuracy of these assembled signals as future work, which we discuss in more detail in Section~\ref{rs:sec:discussion}.}

We provide the parameter settings and versions for each tool as well as the details of the preset parameters in \rs in Supplementary Tables~\ref{rs:tab:parameters} (parameters),~\ref{rs:tab:presets} (details of presets), and~\ref{rs:tab:versions} (versions). We provide the scripts to fully reproduce our results on the GitHub repository at \rsrelease, which contains the corresponding release version of \rs we show in Supplementary Table~\ref{rs:tab:versions}.
We provide the detailed reasoning behind the choice of Flye in order to generate the gold standard in Supplementary Material~\ref{rs:suppsec:gold_standard}.
 
\subsection{Performance and Memory} \label{rs:subsec:perfmemory}

\head{Throughput} Table~\ref{rs:tab:throughput} shows the throughput (i.e., signals processed per second per CPU thread) that \rs reports. Our goal is to estimate the number of CPU threads needed to achieve a throughput faster than the throughput of a single sequencer. Using as few CPU threads as possible is useful to 1)~provide better scalability (i.e., analyzing a larger amount of data with the same computation capabilities) and 2)~reduce the overall computational requirements and corresponding energy consumption (i.e., analyzing the same amount of data with less computational capabilities). The latter is especially essential for resource-constrained devices (e.g., devices with external and limited batteries). The throughput of a single nanopore is around 5,000 signals per second~\cite{oxford_nanopore_technologies_dorado_2024}, and the entire sequencer is usually equipped with 512 nanopores. This means a single sequencer's throughput is around $2{,}560{,}000$ signals per second ($5{,}000 \times 512$)~\cite{magi_nanopore_2018}.
We find that \rs provides an average throughput of \rsavgthr signals per second \textbf{per CPU thread}. This means \rs can achieve an analysis throughput faster than a single sequencer's throughput by using an average of two CPU threads. This shows that the overlapping mechanism of \rs is fast enough for performing real-time overlapping tasks using very few CPU threads.

\begin{table}[htb]
\centering
\caption{\rs Throughput (signals processed per second per CPU thread).}
\resizebox{\columnwidth}{!}{
\begin{tabular}{@{}lrrrrrr@{}}\toprule
                     & D1             & D2           & D3                 & D4           & D5          \\
                     & \emph{E. coli} & \emph{Yeast} & \emph{Green Algae} & \emph{Human} & \emph{E. coli} (R10.4.1)\\\toprule
\textbf{Throughput}  & 2,171,615      & 1,304,910    & 1,117,183          & 1,295,240    & 1,776,225    \\\bottomrule
\end{tabular}

}
\label{rs:tab:throughput}
\end{table}

\head{Computational resources} Table~\ref{rs:tab:resources} shows the computational resources, \revc{and Supplementary Table~\ref{rs:tab:awscost} shows the estimated cloud computing costs to execute each tool for various datasets.} Inside the parentheses provided with \texttt{Dorado + Minimap2} results, we provide the ratio between the reported result and the corresponding \rs result (if higher than $1\times$, \rs is better).

\begin{table}[tbh]
\centering
\caption{Comparison of various tools across different organisms in terms of elapsed time, CPU time, and peak memory usage. The values in parentheses represent the ratio of the result shown in the cell compared to the corresponding result of \rs (values higher than $1\times$ indicate that \rs performs better). We highlight the cells that tools provide a \colorbox{bestresult}{better} and  \colorbox{worseresult}{worse} results with colors.}
\resizebox{\columnwidth}{!}{
\begin{tabular}{@{}llrrr@{}}
\toprule
\textbf{Organism} & \textbf{Tool} & \textbf{Elapsed time} & \textbf{CPU time} & \textbf{Peak} \\
\                 &                & \textbf{(hh:mm:ss)}   & \textbf{(sec)}    & \textbf{Mem. (GB)} \\\toprule
D1 & \textbf{\rs} & 0:52:07 & 184,938 & 7.41 \\
\emph{E. coli} & \textbf{Minimap2 + Dorado CPU (Fast)} & \cellcolor{bestresult}{2:56:45 ($3.39\times$)} & \cellcolor{bestresult}{593,764 ($3.21\times$)} & \cellcolor{bestresult}{36.73 ($4.96\times$)} \\
& \textbf{Minimap2 + Dorado CPU (HAC)} & \cellcolor{bestresult}{13:31:19 ($15.57\times$)} & \cellcolor{bestresult}{1,408,712 ($7.62\times$)} & \cellcolor{bestresult}{62.34 ($8.41\times$)} \\
& \textbf{Minimap2 + Dorado GPU (HAC)} & \cellcolor{worseresult}{0:26:48 ($0.51\times$)} & NA & \cellcolor{bestresult}{26.73 ($3.61\times$)} \\
\midrule
D2 & \textbf{\rs} & 0:01:21 & 3,753 & 6.68 \\
\emph{Yeast} & \textbf{Minimap2 + Dorado CPU (Fast)} & \cellcolor{bestresult}{0:31:11 ($23.10\times$)} & \cellcolor{bestresult}{98,558 ($26.26\times$)} & \cellcolor{bestresult}{146.95 ($22.00\times$)} \\
& \textbf{Minimap2 + Dorado CPU (HAC)} & \cellcolor{bestresult}{2:30:28 ($111.46\times$)} & \cellcolor{bestresult}{548,495 ($146.15\times$)} & \cellcolor{bestresult}{290.85 ($43.54\times$)} \\
& \textbf{Minimap2 + Dorado GPU (HAC)} & \cellcolor{bestresult}{0:02:00 ($1.48\times$)} & NA & \cellcolor{worseresult}{5.98 ($0.90\times$)} \\
\midrule
D3 & \textbf{\rs} & 0:06:20 & 21,769 & 10.87 \\
\emph{Green Algae} & \textbf{Minimap2 + Dorado CPU (Fast)} & \cellcolor{bestresult}{0:30:17 ($4.78\times$)} & \cellcolor{bestresult}{103,102 ($4.74\times$)} & \cellcolor{bestresult}{38.18 ($3.51\times$)} \\
& \textbf{Minimap2 + Dorado CPU (HAC)} & \cellcolor{bestresult}{7:29:20 ($70.95\times$)} & \cellcolor{bestresult}{178,106 ($8.18\times$)} & \cellcolor{bestresult}{20.76 ($1.91\times$)} \\
& \textbf{Minimap2 + Dorado GPU (HAC)} & \cellcolor{worseresult}{0:03:04 ($0.48\times$)} & NA & \cellcolor{worseresult}{10.08 ($0.93\times$)} \\
\midrule
D4 & \textbf{\rs} & 0:45:08 & 136,966 & 6.39 \\
\emph{Human} & \textbf{Minimap2 + Dorado CPU (Fast)} & \cellcolor{bestresult}{5:54:27 ($7.85\times$)} & \cellcolor{bestresult}{1,284,069 ($9.38\times$)} & \cellcolor{bestresult}{113.14 ($17.71\times$)} \\
& \textbf{Minimap2 + Dorado CPU (HAC)} & \cellcolor{bestresult}{18:45:18 ($24.93\times$)} & \cellcolor{bestresult}{4,226,869 ($30.86\times$)} & \cellcolor{bestresult}{130.87 ($20.48\times$)} \\
& \textbf{Minimap2 + Dorado GPU (HAC)} & \cellcolor{worseresult}{0:13:48 ($0.31\times$)} & NA & \cellcolor{bestresult}{17.16 ($2.69\times$)} \\
\midrule
D5 & \textbf{\rs} & 16:04:39 & 2,672,608 & 112.31 \\
\emph{E. coli} & \textbf{Minimap2 + Dorado CPU (Fast)} & \cellcolor{bestresult}{17:16:52 ($1.07\times$)} & \cellcolor{bestresult}{3,478,136 ($1.30\times$)} & \cellcolor{worseresult}{103.1 ($0.92\times$)} \\
(R10.4.1) & \textbf{Minimap2 + Dorado CPU (HAC)} & \cellcolor{bestresult}{135:05:47 ($8.40\times$)} & \cellcolor{bestresult}{19,051,635 ($7.13\times$)} & \cellcolor{worseresult}{103.1 ($0.92\times$)} \\
& \textbf{Minimap2 + Dorado GPU (HAC)} & \cellcolor{worseresult}{4:59:49 ($0.31\times$)} & NA & \cellcolor{worseresult}{103.1 ($0.92\times$)} \\
\bottomrule
\end{tabular}

}
\label{rs:tab:resources}
\end{table}

We make three key observations.
First, \rs provides a substantial speedup and lower peak memory usage compared to Dorado's fast model on a CPU followed by minimap2 (\texttt{Dorado CPU (Fast) + Minimap2}) by, on average, \rsavgelfcpu (elapsed time), \rsavgcpufcpu (CPU time), and \rsavgpeakfcpu (peak memory usage). When using Dorado's HAC model on a CPU, \rs provides even better results on average by \rsavgelhcpu (elapsed time), \rsavgcpuhcpu (CPU time), and \rsavgpeakhcpu (peak memory usage), since running the HAC model is computationally more costly than running the fast model.

\revb{Second, GPU-based basecalling followed by minimap2 takes \rsavgelhgpu of the time that \rs takes while requiring higher peak memory usage by \rsavgpeakhgpu compared to \rs.} Although comparing the results between CPUs and GPUs is not ideal due to the massive parallelism that GPUs provide compared to CPUs, \revb{\rs still provides comparable performance given the limited parallelism it uses on CPUs (i.e., 64 threads) compared to the massive parallelism that GPUs provide (e.g., a few tens of thousands of threads).} This shows that \rs can provide even faster results when accelerated with GPUs based on its comparison when basecalling is done using CPUs and GPUs, which we leave as future work.

\revc{Third, in most cases where \rs is slower than GPU-based configurations, \rs still provides a cheaper solution when using cloud computing services. This is mainly because GPU usage in the cloud is associated with substantially higher monetary costs than using a CPU-only solution under today's pricing scheme. We find that \rs provides the cheapest solution for identifying overlaps in most cases, even when it is slower than GPU-based configurations.}

We conclude that \rs can perform read overlapping with higher throughput that can be useful when focusing on real-time analysis and better computational resources than CPU-based basecalling followed by minimap2. We find that \rs's speed is mainly dependent on the amount of bases stored in the hash table, as the speed decreases with increasing the number of locations that need to be analyzed in the index per read. To resolve this scalability issue, future work should focus on designing indexing and filtering methods that provide a limitation on the volume of signals stored in the index and processed during the overlapping step. These results can also be useful when basecalling raw signals using GPUs to reduce the computational overhead that GPU-based basecalling requires, which we discuss in Section~\ref{rs:sec:discussion}.

\subsection{All-vs-All Overlapping Statistics} \label{rs:subsec:accuracy}
Table~\ref{rs:tab:accuracy} shows the all-vs-all overlapping statistics between \rs and minimap2. We make two key observations.
First, on average, \rsavgshared of the overlap pairs generated by \rs are shared with the overlap pairs that minimap2 generates. This shows that a large fraction of overlapping pairs does not require basecalling to generate identical overlapping information identified by minimap2 after basecalling.
Second, although \rs can find a substantial amount of overlap pairs identical to the overlaps minimap2 finds, there are still overlapping pairs unique to \rs (\rsavgru on average) and minimap2 (\rsavgmmu on average). These differences are likely due to differences in 1)~certain parameters (e.g., chaining scores) and 2)~increased noise inherent in raw signals compared to basecalled sequences, which can become more pronounced in genomes with greater size, complexity, or repetitiveness. \revc{These factors overall result in fewer overlaps found by \rs. Although certain parameters can be configured (e.g., minimum chaining scores) to find more overlaps, this strategy usually results in finding a substantially larger number of erroneous overlaps in \rs. Designing robust mechanisms that can handle noise better and more accurate hashing mechanisms with fewer collisions can provide better trade-offs.} \rev{Although maximizing the shared overlaps can provide insights regarding the accuracy of overlapping information that \rs finds, \textbf{it is not necessary to provide near-identical shared overlap statistics} for certain use cases such as constructing \emph{de novo} assemblies~\cite{vaser_time-_2021}.} Instead, contiguous and more accurate assemblies can still be constructed using a smaller but useful portion of shared overlapping pairs~\cite{vaser_time-_2021}. We conclude that \rs provides a mechanism that shares a large portion of overlaps with minimap2, while the shared portions decrease due to the differences in parameters and the noise in raw signal datasets.

\begin{table}[tbh]
\centering
\caption{All-vs-all Overlapping Statistics. Percentages show the overlapping pairs that are 1)~unique to \rs, 2)~unique to minimap2, and 3)~reported in both tools (Shared Overlaps).}
\resizebox{\columnwidth}{!}{
\begin{tabular}{@{}clrrr@{}}\toprule
& \textbf{Organism}      & \textbf{Unique to}     & \textbf{Unique to} & \textbf{Shared} \\
&                        & \textbf{\rs ($\%$)}  & \textbf{Minimap2 ($\%$)} & \textbf{Overlaps ($\%$)} \\\toprule
D1 & \emph{E. coli}      & 8.33 & 50.62 & 41.05 \\\midrule
D2 & \emph{Yeast}        & 26.54 & 28.62 & 44.84 \\\midrule
D3 & \emph{Green Algae}  & 3.76 & 78.64 & 17.61 \\\midrule
D4 & \emph{Human}        & 25.65 & 62.62 & 11.73\\\midrule
D5 & \emph{E. coli} (R10.4.1) & 0.06 & 79.69 & 20.25 \\\bottomrule
\end{tabular}

}
\label{rs:tab:accuracy}
\end{table}

\subsection{\rev{Contiguity and Estimated Accuracy of the Assembly Graphs}} \label{rs:subsec:assembly}
To evaluate the impact of overlaps that \rs and minimap2 find, we construct \emph{de novo} assembly \rev{graphs} from these overlaps by using miniasm off-the-shelf. Table~\ref{rs:tab:assembly} shows the contiguity statistics and estimated accuracy of these \emph{de novo} assembly \rev{graphs} \revb{(please see Section~\ref{rs:subsec:evaluation} for our explanation and discussion related to using these accuracy ratios \emph{only} for comparison purposes, not for calculating the actual assembly identities)}. \rev{We note that the assembly graphs constructed using miniasm provide the connection information used to assemble unitigs, as well as the connections between unitigs that make up subgraph components in the assembly graph. Since we use the miniasm assembler without any modifications, we cannot generate the assembled sequences from raw signals as miniasm supports generating assembled sequences from basecalled sequencing reads. \revb{This mainly prevents us from evaluating the actual identity of these assemblies from raw signals, which we further discuss in Section~\ref{rs:sec:discussion}.}}

\begin{table}[tbh]
\centering
\caption{Assembly Statistics. Chained Read Percentage provides our estimate in terms of assembly graph accuracy. Other columns provide our evaluations in terms of assembly contiguity. For the accuracy, we highlight the cells that tools provide a \colorbox{bestresult}{better} and  \colorbox{worseresult}{worse} results with colors.}
\resizebox{\columnwidth}{!}{
\begin{tabular}{@{}llrrrrrr@{}}
\toprule
\textbf{Dataset}   & \textbf{Tool} & \textbf{Chained Read}         & \textbf{Largest}                    & \textbf{N50}                       & \textbf{auN}                       & \textbf{Longest}                   & \textbf{Unitig} \\
                   &               & \textbf{Percentage (\%)}      & \textbf{Comp. (bp)}                 & \textbf{(bp)}                      & \textbf{(bp)}                      & \textbf{Unitig (bp)}               & \textbf{Count}  \\\toprule
D1                 & \rs           & \cellcolor{bestresult}{34.0}  & \cellcolor{worseresult}{4,841,669}  & \cellcolor{worseresult}{1,535,079} & \cellcolor{worseresult}{1,187,229} & \cellcolor{worseresult}{2,347,310} & \cellcolor{worseresult}{32} \\
\emph{E. coli}     & minimap2      & \cellcolor{worseresult}{24.3} & \cellcolor{bestresult}{5,207,206}   & \cellcolor{bestresult}{5,204,754}  & \cellcolor{bestresult}{5,194,495}  & \cellcolor{bestresult}{5,207,206}  & \cellcolor{bestresult}{4}   \\
                   & Gold standard & NA                            & 5,235,343                           & 5,235,343                          & 5,235,343                          & 5,235,343                          & 1 \\\midrule
D2                 & \rs           & \cellcolor{bestresult}{43.5}  & \cellcolor{worseresult}{362,050}    & \cellcolor{worseresult}{41,118}    & \cellcolor{worseresult}{48,106}    & \cellcolor{bestresult}{161,883}   & \cellcolor{worseresult}{396} \\
\emph{Yeast}       & minimap2      & \cellcolor{worseresult}{43.2} & \cellcolor{bestresult}{1,611,876}   & \cellcolor{bestresult}{134,050}    & \cellcolor{bestresult}{137,172}    & \cellcolor{worseresult}{64,054}     & \cellcolor{bestresult}{282}  \\
                   & Gold standard & NA                            & 11,835,059                          & 640,934                            & 623,210                            & 1,073,346                          & 68 \\\midrule
D3                 & \rs           & \cellcolor{worseresult}{41.3} & \cellcolor{bestresult}{448,422}     & \cellcolor{bestresult}{93,111}     & \cellcolor{bestresult}{107,914}    & \cellcolor{bestresult}{252,038}    & \cellcolor{bestresult}{50} \\
\emph{Green Algae} & minimap2      & \cellcolor{bestresult}{53.2}  & \cellcolor{worseresult}{198,709}    & \cellcolor{worseresult}{63,310}    & \cellcolor{worseresult}{91,360}    & \cellcolor{worseresult}{198,709}   & \cellcolor{worseresult}{55} \\
                   & Gold standard & NA                            & 2,255,807                           & 452,774                            & 538,136                            & 1,667,975                          & 420 \\\midrule
D4                 & \rs           & \cellcolor{worseresult}{61.2} & \cellcolor{bestresult}{183,402}     & \cellcolor{bestresult}{47,397}     & \cellcolor{bestresult}{66,574}     & \cellcolor{bestresult}{183,402}    & \cellcolor{bestresult}{48} \\
\emph{Human}       & minimap2      & \cellcolor{bestresult}{72.7}  & \cellcolor{worseresult}{81,017}     & \cellcolor{worseresult}{19,459}    & \cellcolor{worseresult}{26,925}    & \cellcolor{worseresult}{81,017}    & \cellcolor{worseresult}{64} \\
                   & Gold standard & NA                            & 367,305                             & 19,329                             & 29,697                             & 150,470                            & 592 \\\midrule
D5                 & \rs           & \cellcolor{bestresult}{34.6}  & \cellcolor{worseresult}{22,921,373} & \cellcolor{bestresult}{27,460}     & \cellcolor{worseresult}{43,505}    & \cellcolor{worseresult}{410,905}   & \cellcolor{worseresult}{2,194} \\
\emph{E. coli}     & minimap2      & \cellcolor{worseresult}{16.3} & \cellcolor{bestresult}{27,250,402}  & \cellcolor{worseresult}{24,111}    & \cellcolor{bestresult}{814,807}    & \cellcolor{bestresult}{5,117,413}  & \cellcolor{bestresult}{2,888} \\
(R10.4.1)          & Gold standard & NA                            & 5,230,610                           & 5,230,610                          & 5,230,610                          & 5,230,610                          & 1 \\\bottomrule
\end{tabular}

}
\label{rs:tab:assembly}
\end{table}

\rev{First, we find that we can construct long unitigs from the raw signal overlaps that \rs finds. The unitigs we can construct from these raw signal overlaps are substantially longer than the average read length of their corresponding datasets (e.g., the longest unitig length in D1 is $413.25\times$ longer than the average read length of the D1 dataset). \textbf{\rs is the first work that enables \emph{de novo} assembly construction directly from raw signal overlaps without basecalling}, which has several implications and can enable future work, as we discuss in Section~\ref{rs:sec:discussion}.}

Second, we observe that the unitigs generated from \rs overlaps are usually less contiguous compared to those generated from minimap2 overlaps, based on all the metrics we show in Table~\ref{rs:tab:assembly}. \revc{We believe this is mainly inline with the all-vs-all overlapping statistics where minimap2 finds more overlaps than \rs, leading to longer assemblies.} Compared to the gold standard assemblies generated using highly accurate basecalled reads and a state-of-the-art assembler, we find that \rs can still achieve a significant portion of the assembly contiguity, especially for the D1 dataset (\emph{E.~coli}). For example, \rs constructs unitigs with the longest unitig length of 2.3 million bases, which is almost half the length of the \emph{E.~coli} genome, whereas the gold standard assembly produces a single unitig covering the entire genome. \textbf{This indicates that \rs can achieve substantial contiguity relative to the gold standard without basecalling.}

\revb{Third, in terms of the accuracy estimate of the assembly graphs (i.e., \textit{Chained Read Percentage}), we find that a larger percentage of \emph{unitig reads} generated by the \rs overlaps appear in close proximity when these reads are mapped to their corresponding reference genome compared to the unitig reads generated by the minimap2 overlaps, \revc{except for two cases (i.e., D3 and D4).} Although this observation by itself does not show that the \rs overlaps can lead to assemblies with higher identity, it provides the accuracy estimate that these assemblies share similar accuracy to those generated by the minimap2 overlaps.} \revc{For the cases where \rs provides a lower chained read percentage, \rs leads to better contiguity. These datasets are from larger genomes with lower sequencing depth. We believe that spurious overlaps are more common at lower coverage when using raw signals, leading to contiguous but erroneous assemblies.}

We conclude that \rs generates useful overlapping information, enabling the construction of assemblies directly from raw signals. \revb{The assembly graphs generated by the \rs overlaps mainly provide comparable contiguity and accuracy compared to those generated by minimap2.} \rev{We discuss the potential next steps to enable evaluating the accuracy of these assemblies in Section~\ref{rs:sec:discussion}.}

\section{Discussion and Future Work} \label{rs:sec:discussion}
\head{Limitations} Our evaluation demonstrates that \rs achieves high throughput when finding overlaps between raw signal pairs, making it a viable candidate for real-time analysis. However, there are still two main challenges to fully utilize real-time \emph{de novo} assembly construction during sequencing.
First, the hash-based index should be constructed and updated dynamically in real-time while sequencing is in progress. Although \rs provides the mechanisms for storing multiple hash tables that can be constructed for each chunk of raw signals generated in real-time, it is not computationally feasible to dynamically update the index for all sequenced signals as the memory and computational burden increase with each hash table generated. Such an approach requires a decision-making mechanism to dynamically stop updating the index after sequencing a certain amount of signals. The stopping mechanism should be accurate enough to ensure that the current state of the index provides sufficient information to find the overlaps between the already sequenced signals and the new signals generated after the index construction.
Second, the assembly graph should be constructed and updated dynamically while the new overlap information is generated in real time. This is challenging as the intermediate steps for generating the unitigs (e.g., the transitive reduction step~\cite{myers_fragment_2005}) in assembly graphs may not work optimally without the full overlap information, as it is likely to remove graph connections that can be useful with new overlap information. It might be feasible to use graph structures that are more suitable for streaming data generation, such as de Bruijn graphs, for assembly construction purposes~\cite{el-metwally_lightassembler_2016, rozov_faucet_2018, scott_streaming_2022}.

\rs can find overlaps only between reads coming from the same strand, as it lacks the capability to construct the reverse-complemented version of the signals to identify matches on the other strand. \revc{\rs cannot detect reverse-complemented strands, since this requires modifying the original raw signal to generate its reverse-complemented version. Our preliminary results show that simple one-to-one mapping strategies are not practical, potentially due to noise.} Lacking reverse complemented signals potentially leads to gaps in the assembly and construction of the unitigs from both strands. Designing a mechanism that can reverse complement raw signals without basecalling is future work.

\rev{\head{Challenges for constructing and evaluating \emph{de novo} assemblies} Although the main focus of our work is enabling all-vs-all overlapping between raw signals, we identify \emph{de novo} assembly construction as a natural use case that can utilize the all-vs-all overlap information from raw signals that \rs finds. To show that the overlaps from \rs can lead to unitigs that are much longer than average read lengths, we construct and evaluate the \emph{de novo} assembly graphs from the \rs overlaps, as discussed in Section~\ref{rs:subsec:assembly}. However, we identify two key challenges that limit the scope of our evaluation of these \emph{de novo} assemblies, which we leave as future work.

First, we cannot generate the sequence of assembled signals from the assembly graphs (i.e., spelling the sequence from the assembly graph). This is because we use the miniasm assembler to construct these assembly graphs from the raw signal overlaps. Although miniasm can use the overlap position information to construct the assembly graph, it cannot output the sequence of assemblies, as it is mainly designed for basecalled sequences. This is challenging because each raw signal contains metadata used in basecalling mechanisms (e.g., nanopore-specific information such as the baseline current levels). Carrying the metadata information effectively from many overlapping signals is essential to enable further analysis of these assembled signals.

Second, to evaluate the identity of the constructed assemblies, these assemblies need to be aligned to a ground truth assembly. This can be done by either 1)~basecalling the assembled signal or 2)~aligning the assembled signal using dynamic time warping (DTW)~\cite{lindegger_rawalign_2023} to the ground truth assembly. These approaches require constructing the assembled signal (i.e., the first challenge mentioned above) and designing a new class of basecallers to directly basecall assembled signals, which we discuss as future work below. Although the missing identity evaluations of these assemblies can be a major limitation of our work, we believe the assemblies that are constructed from the \rs overlaps are likely to be close to the assembly accuracy that can be achieved by using the minimap2 overlaps \revb{based on our estimated accuracy evaluation.}}

\head{New Directions for Future Work}
We identify several new directions for future work. First, integrating the overlapping information with basecalling can provide accuracy benefits for basecallers \revc{(e.g., basecalling the overlapped region only once or designing basecallers such that they collectively take overlapping region for more accurate basecalling).} Existing basecallers are designed to basecall individual reads without such overlapping information. Such a combined approach can provide additional useful features for the underlying machine learning models in basecallers to improve the basecalling accuracy.
Second, \emph{de novo} assembly construction enables identifying the reads that do not provide useful information for assembly if they are fully contained by other overlapping read pairs~\cite{li_minimap_2016}. Identifying these potentially useless reads early on provides the opportunity to avoid basecalling them, which can improve the overall execution time of basecalling by filtering out such reads.
Third, \emph{de novo} assembly construction from raw signals provides new opportunities to design new basecallers that can basecall the assembled signals. Such a basecaller has the potential to substantially improve the performance, as it enables basecalling fewer long unitigs instead of many shorter reads.
\revc{Fourth, existing assemblies generated from basecalled sequences can leverage the raw signal information to resolve complex regions and perform phasing by utilizing the rich information in signals that may resolve certain repeats.}
\revb{Fifth, more accurate and robust segmentation approaches for signal processing~\cite{bakic_campolina_2025} can lead to better overlapping and assembly construction, which requires further investigation for this particular use case.}

While \rs enables new directions in raw signal analysis, particularly for \emph{de novo} assembly, several challenges and opportunities for future work remain. Addressing these challenges has the potential to enable new use cases and applications in raw signal and basecalled sequence analyses.

\section{Conclusion} \label{rs:sec:conclusion}
We introduce \rs, the \emph{first} mechanism that can find overlaps between two sets of raw nanopore signals without translating them to bases.
We find that \rs can 1)~find overlaps while meeting the real-time requirements with an average throughput
of \rsavgthr signals/sec per CPU thread, 2)~reduce the overall time needed for finding overlaps (on average by \rsavgelfcpu and up to by \rsmaxelfcpu) and peak memory usage (on average by \rsavgpeakfcpu and up to by \rsmaxpeakfcpu) compared to the time and memory needed to run state-of-the-art read mapper (minimap2) combined with the Dorado basecaller running on a CPU with its fastest model, 3)~share a large portion of overlapping pairs with minimap2 (\rsavgshared on average), and 4)~construct long assemblies from these useful overlaps. We find that we can construct assemblies of half the length of the entire \emph{E. coli} genome (i.e., of length 2.3 million bases) directly from raw signal overlaps without basecalling. Finding overlapping pairs from raw signals is critical for enabling new directions that have not been explored before for raw signal analysis, such as \emph{de novo} assembly construction from overlaps that we explore in this work. We discuss many other new directions that can be enabled by finding overlaps and constructing \emph{de novo} assemblies.

We hope and believe that \rs enables future work in at least two key directions. First, we aim to fully perform end-to-end genome analysis without basecalling. This can be achieved by further improving tools such as \rs to construct \emph{de novo} assemblies directly from raw signals such that these assemblies are consistently better than those generated from basecalled sequences in all cases. Second, we should rethink how we train and use modern neural network-based basecallers by integrating additional useful information (e.g., overlaps or assemblies of signals) generated by \rs into these basecallers.

\section*{Acknowledgments}
C.F. acknowledges gift funding provided by AMD. SAFARI Research Group acknowledges the generous gift funding provided by industrial partners (especially by Google, Intel, Microsoft, VMware), which has been instrumental in enabling the 20+ year-long research SAFARI Research Group has been conducting on accelerating genome analysis. SAFARI Research Group is also partially supported by the Semiconductor Research Corporation (SRC), the European Union’s Horizon programme for research and innovation [101047160 - BioPIM] and the Swiss National Science Foundation (SNSF) [200021\_213084].
M.M. has been supported by the Max Planck ETH Center for Learning Systems and by SNSF Project Grant \#200550 to A.K.
S.G. was funded through SNSF Project Grant \#200550 to A.K.

\bibliographystyle{IEEEtran}
{\bibliography{main}}
\setstretch{1}
\onecolumn
\newcommand{\supptitle}[1]{%
  \begingroup
    \fontsize{17pt}{19pt}\selectfont
    \bfseries
    #1\par
  \endgroup
}
\setcounter{secnumdepth}{3}
\clearpage
\begin{center}
\supptitle{Supplementary Material for\\\rsltitle}
\end{center}
\setcounter{section}{0}
\setcounter{equation}{0}
\setcounter{figure}{0}
\setcounter{table}{0}
\setcounter{page}{1}
\makeatletter
\renewcommand{\theequation}{S\arabic{equation}}
\renewcommand{\thetable}{S\arabic{table}}
\renewcommand{\thefigure}{S\arabic{figure}}
\renewcommand{\thesection}{\Alph{section}}
\renewcommand{\thesubsection}{\thesection.\arabic{subsection}}
\renewcommand{\thesubsubsection}{\thesubsection.\arabic{subsubsection}}

\newcommand{\TextUnderscore}{\rule{.4em}{.4pt}}
\section{\rht Overview}\label{rs:suppsec:rht_overview}

\rs builds improvements over \rht~\citesupp{supp_firtina_rawhash_2023, supp_firtina_rawhash2_2024}, a mechanism that provides a hash-based similarity identification between a raw signal and a reference genome. We show the overview of \rht in Supplementary Figure~\ref{rs:suppfig:rh2_overview}. \rht has four key steps.

First, to generate sequences of signals that can be compared to each other, \rht generates signals of k-mers, called \emph{events}, from both a reference genome and raw signals. To generate events from reference genomes, it uses a lookup table, called \emph{k-mer model}, that provides the expected signal value (i.e., event value) as a floating value for each possible k-mer where k is usually 6 or 9, depending on the flow cell version. To identify events (i.e., k-mers) in raw signals, \rht performs a segmentation technique to detect the abrupt changes in signals, which enables identifying the regions in signals generated when sequencing a particular k-mer. \rht uses the average value of signals within the same region as an event value after identifying outliers within the region. Due to the variations, oversegmentation issues used in the classical segmentation algorithms (e.g., t-test), and noise in nanopore sequencing, event values generated from the same k-mer can slightly differ from each other, making it challenging to directly match the event values to each other to identify matching k-mers between a reference genome and raw signals. \revb{These variations are usually handled by performing homopolymer compression (HPC) type strategies, where consecutive signals with similar values are collapsed into a single signal~\citesupp{supp_zhang_real-time_2021, supp_firtina_rawhash_2023, supp_firtina_rawhash2_2024,supp_shivakumar_sigmoni_2024}.}

Second, to further mitigate this noise issue in events, \rht quantizes the event values such that slightly different event values can be quantized into the same value (i.e., bucketing) to enable direct matching of quantized event values between a reference genome and raw signals. To enable an accurate quantization, \rht identifies the range of values that are assigned to the same quantized value dynamically according to the nanopore model.

Third, to reduce the number of potential matches without reducing accuracy, \rht concatenates the quantized event values of consecutive events (i.e., consecutive k-mers) and generates a hash value from these concatenated values.

Fourth, for the reference genome, these hash values are stored in a hash table along with their position information, which is usually known as the indexing step in read mapping. \rht uses the hash values of raw signals to query the previously constructed hash table to identify matching hash values, known as \emph{seed hits}, between a reference genome and a raw signal. Seed hits are used to identify chains of seed matches within proximity using the chaining algorithm proposed in minimap2~\citesupp{supp_li_minimap2_2018}. \rht identifies the best chain among many chains based on a weighted decision strategy that takes several metrics into account, such as chaining score and mapping quality. \rht uses the best chain as the best mapping for a read and reports a single mapping per read.

\begin{figure}[tbh]
  \centering
  \includegraphics[width=0.7\columnwidth]{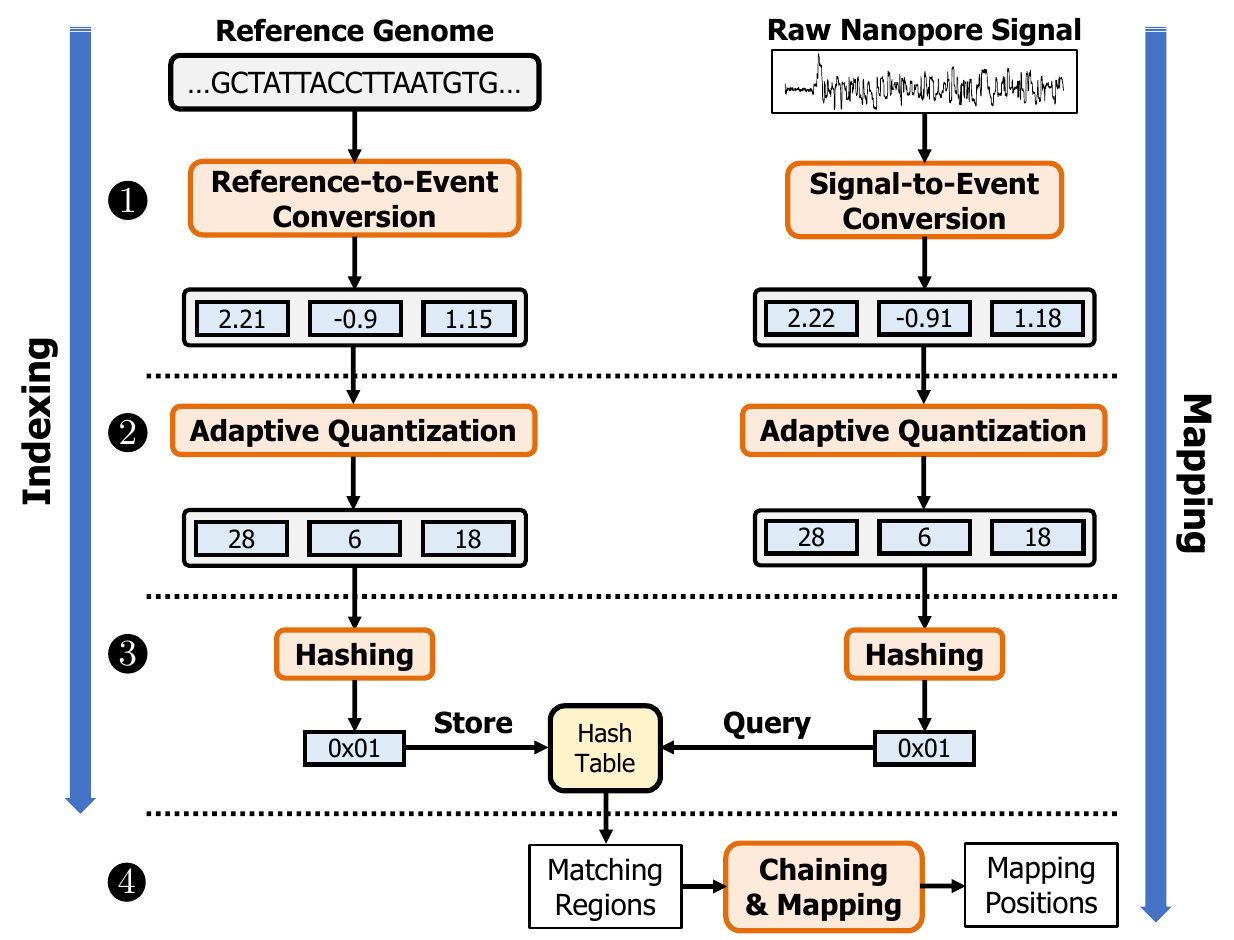}
  \caption{Overview of \rht.}
  \label{rs:suppfig:rh2_overview}
\end{figure}

\clearpage
\revc{
\section{Estimating the AWS Cloud Computing Costs}
\label{rs:suppsec:aws_pricing}

Supplementary Table~\ref{rs:tab:awscost} shows the estimated cloud computing prices when using the corresponding instances provided by Amazon Web Services (AWS). We estimate AWS compute costs for each tool by selecting the cheapest EC2 instance in the \texttt{us-east-1} region such that 1)~the architecture is \texttt{x86\_64}, 2)~the vCPU count is fixed to 64 to match our experimental configuration, and 3)~the \textbf{memory} is at least the measured peak memory reported in Table~\ref{rs:tab:resources}. For GPU basecalling, we additionally require a single GPU with 48~GB VRAM.

To compute a \emph{compute-only} monetary cost estimate, we multiply the instance on-demand hourly rate as provided by AWS\footnote{We use the online calculator at \url{https://calculator.aws/\#/createCalculator/ec2-enhancement}} by the measured wall-clock elapsed time. We report the ratio of each tool's estimated cost relative to \rs on the same dataset (values in parentheses). We highlight rows where \rs is \colorbox{bestresult}{cheaper} or \colorbox{worseresult}{more expensive}. This estimate excludes storage and data transfer charges, and prices are specific to the selected region and the retrieval date.

\begin{table}[tbh] 
\centering
\caption{Estimated monetary costs when using AWS cloud computing to execute each workload.}
\begin{tabular}{@{}lllrll@{}}
\toprule
\textbf{Organism} & \textbf{Tool} & \textbf{AWS instance} & \textbf{Rate} & \textbf{Elapsed} & \textbf{Est. cost}\\
 &  & (64 vCPU, x86\_64) & (USD/hr) & (hh:mm:ss) & (USD)\\
\toprule
D1 & \textbf{\rs} & \texttt{c6a.16xlarge} & 2.4480 & 0:52:07 & 2.13 \\
\emph{E. coli} & \textbf{Dorado CPU (Fast) + Minimap2} & \texttt{c6a.16xlarge} & 2.4480 & 2:56:45 & 7.21 \cellcolor{bestresult}{($3.39\times$)} \\
 & \textbf{Dorado CPU (HAC) + Minimap2} & \texttt{c6a.16xlarge} & 2.4480 & 13:31:19 & 33.10 \cellcolor{bestresult}{($15.57\times$)} \\
 & \textbf{Dorado GPU (HAC) + Minimap2} & \texttt{g6e.16xlarge} & 7.5772 & 0:26:48 & 3.38 \cellcolor{bestresult}{($1.59\times$)} \\
\midrule
D2 & \textbf{\rs} & \texttt{c6a.16xlarge} & 2.4480 & 0:01:21 & 0.06 \\
\emph{Yeast} & \textbf{Dorado CPU (Fast) + Minimap2} & \texttt{m5a.16xlarge} & 2.7520 & 0:31:11 & 1.43 \cellcolor{bestresult}{($25.97\times$)} \\
 & \textbf{Dorado CPU (HAC) + Minimap2} & \texttt{r6a.16xlarge} & 3.6288 & 2:30:28 & 9.10 \cellcolor{bestresult}{($165.22\times$)} \\
 & \textbf{Dorado GPU (HAC) + Minimap2} & \texttt{g6e.16xlarge} & 7.5772 & 0:02:00 & 0.25 \cellcolor{bestresult}{($4.59\times$)} \\
\midrule
D3 & \textbf{\rs} & \texttt{c6a.16xlarge} & 2.4480 & 0:06:20 & 0.26 \\
\emph{Green Algae} & \textbf{Dorado CPU (Fast) + Minimap2} & \texttt{c6a.16xlarge} & 2.4480 & 0:30:17 & 1.24 \cellcolor{bestresult}{($4.78\times$)} \\
 & \textbf{Dorado CPU (HAC) + Minimap2} & \texttt{c6a.16xlarge} & 2.4480 & 7:29:20 & 18.33 \cellcolor{bestresult}{($70.95\times$)} \\
 & \textbf{Dorado GPU (HAC) + Minimap2} & \texttt{g6e.16xlarge} & 7.5772 & 0:03:04 & 0.39 \cellcolor{bestresult}{($1.50\times$)} \\
\midrule
D4 & \textbf{\rs} & \texttt{c6a.16xlarge} & 2.4480 & 0:45:08 & 1.84 \\
\emph{Human} & \textbf{Minimap2 + Dorado CPU (Fast)} & \texttt{c6a.16xlarge} & 2.4480 & 5:54:27 & 14.46 \cellcolor{bestresult}{($7.85\times$)} \\
 & \textbf{Minimap2 + Dorado CPU (HAC)} & \texttt{m5a.16xlarge} & 2.7520 & 18:45:18 & 51.61 \cellcolor{bestresult}{($28.03\times$)} \\
 & \textbf{Minimap2 + Dorado GPU (HAC)} & \texttt{g6e.16xlarge} & 7.5772 & 0:13:48 & 1.74 \cellcolor{worseresult}{($0.95\times$)} \\
\midrule
D5 & \textbf{\rs} & \texttt{c6a.16xlarge} & 2.4480 & 16:04:39 & 39.36 \\
\emph{E. coli} (R10.4.1) & \textbf{Dorado CPU (Fast) + Minimap2} & \texttt{c6a.16xlarge} & 2.4480 & 17:16:52 & 42.30 \cellcolor{bestresult}{($1.07\times$)} \\
 & \textbf{Dorado CPU (HAC) + Minimap2} & \texttt{c6a.16xlarge} & 2.4480 & 135:05:47 & 330.72 \cellcolor{bestresult}{($8.40\times$)} \\
 & \textbf{Dorado GPU (HAC) + Minimap2} & \texttt{g6e.16xlarge} & 7.5772 & 4:59:49 & 37.86 \cellcolor{worseresult}{($0.96\times$)} \\
\bottomrule
\end{tabular}

\label{rs:tab:awscost}
\end{table}

}
\clearpage
\section{Generating a Gold Standard Assembly}
\label{rs:suppsec:gold_standard}
To generate a gold standard assembly using R9.4 Simplex reads, we explore two approaches. 

First, we use the raw basecalled sequences directly with a set of state-of-the-art assemblers identified from recent benchmarks~\citesupp{Sun2021, Wick2021, Cosma2022, Yu2024}, namely Hifiasm~\citesupp{Cheng2021}, Verkko~\citesupp{Rautiainen2023}, LJA~\citesupp{Bankevich2022}, HiCanu~\citesupp{Nurk2020}, and Flye~\citesupp{Kolmogorov2019}.

Second, we apply error correction to the reads using the HERRO tool~\citesupp{Stanojevic2024} developed by Oxford Nanopore Technologies (ONT), which is also integrated into the latest versions of Dorado as \textit{dorado correct}. HERRO corrects erroneous R9.4 and R10.4 data, enabling their use as a replacement for accurate PacBio HiFi reads required by state-of-the-art hybrid assembly approaches. Supplementary Table~\ref{rs:tab:coverage} shows the estimated sequencing depth of coverage before and after the HERRO correction for each dataset.

For the high-coverage D1 \emph{E.~coli} dataset, Hifiasm outputs a fragmented assembly, while Verkko requires extensive parameter tuning (specifically in terms of coverage, \texttt{---unitig-abundance}, and \texttt{---base-k}) to achieve desirable contiguity and completeness. LJA produces an almost perfect assembly when a minimum length filter of 30kbp is applied to the corrected reads, as suggested by the HERRO paper. Flye performs comparably to LJA without the need for error correction. \revb{It should be noted that the golden standard assembly for D5 \emph{E.~coli} dataset was limited in terms of assembly coverage, using \texttt{---asm-coverage 50}, as Flye raised an error regarding an otherwise too high coverage level.}

For the other datasets, with or without error correction, most assemblers either fail to produce assemblies or generate suboptimal results, except for Flye. We attribute this to their sensitivity to inaccuracies in the uncorrected reads and the reduced coverage after correction. Supplementary Table~\ref{rs:tab:coverage} shows how the coverage levels change after correction for each dataset. Notably, Flye works well (and sometimes even slightly better) with uncorrected reads, highlighting its robustness to noisy ONT reads.

Based on these observations, we select Flye as the assembler to generate the gold standard assemblies in our evaluations, given its ability to handle noisy ONT reads without the need for error correction and its consistent performance across different datasets. Therefore, we use Flye to construct the gold standard assemblies from the basecalled reads in our study.

\begin{table}[tbh] 
\centering
\caption{Coverage levels before and after error correction for each dataset.}
\begin{tabular}{@{}lcc@{}}\toprule
\textbf{Dataset} & \textbf{Coverage Before Correction} & \textbf{Coverage After Correction} \\\midrule
D1 \emph{E. coli} & 445$\times$ & 240$\times$ \\
D2 \emph{Yeast} & 32$\times$ & 12$\times$ \\
D3 \emph{Green algae} & 5.6$\times$ & 3.7$\times$ \\
D4 \emph{Human} & 0.6$\times$ & 0.002$\times$ \\
D5 \emph{E. coli} (R10.4.1) & \revb{1796.9}$\times$ & \revb{338}$\times$ \\\bottomrule
\end{tabular}

\label{rs:tab:coverage}
\end{table}

\clearpage
\section{Configuration} \label{rs:sec:configuration}
\subsection{Parameters} \label{rs:subsec:parameters}

In Supplementary Table~\ref{rs:tab:parameters}, we show the parameters of each tool for each dataset. In Supplementary Table~\ref{rs:tab:presets}, we show the details of the preset values that \rs sets in Supplementary Table~\ref{rs:tab:parameters}. For minimap2~\citesupp{supp_li_minimap2_2018}, we use the same parameter setting for all datasets. For miniasm~\citesupp{supp_li_minimap_2016}, we use the default parameter settings for all datasets.

\begin{table}[tbh]
\centering
\caption{Parameters we use in our evaluation for each tool and dataset in mapping.}
\begin{tabular}{@{}lccccc@{}}\toprule
\textbf{Tool} & \textbf{D1 \emph{E. coli}} & \textbf{D2 \emph{Yeast}} & \textbf{D3 \emph{Green Algae}} & \textbf{D4 \emph{Human}} & \textbf{D5 \emph{E. coli (R10.4.1)}} \\\midrule
\rs    & -x ava -t 64 & -x ava -t 64 & -x ava -t 64 & -x ava -t 64 & -x ava --r10 -t 64 \\\midrule
Minimap2          & \multicolumn{5}{c}{-x ava-ont --for-only -t 64}\\\midrule
Dorado CPU (Fast) & \multicolumn{5}{c}{basecaller -x cpu fast}\\\midrule
Dorado CPU (HAC) & \multicolumn{5}{c}{basecaller -x cpu hac}\\\midrule
Dorado GPU (HAC) & \multicolumn{5}{c}{basecaller hac}\\\midrule
\end{tabular}

\label{rs:tab:parameters}
\end{table}

\begin{table}[tbh]
\centering
\caption{Corresponding parameters of presets (-x) in \rs.}
\resizebox{\linewidth}{!}{
\begin{tabular}{@{}lcc@{}}\toprule
\textbf{Preset} & \textbf{Corresponding parameters} & Usage \\\midrule
ava-viral  & -e 6 -q 4 -w 0 --sig-diff 0.45 --fine-range 0.4 --min-score 20 --min-score2 30 --min-anchors 5 & Viral genomes\\
& --min-mapq 5 --bw 1000 --max-target-gap 2500 --max-query-gap 2500 --chain-gap-scale 1.2 --chain-skip-scale 0.3 &\\\midrule
ava  & -e 8 -q 4 -w 3 --sig-diff 0.45 --fine-range 0.4 --min-score 40 --min-score2 75 & Default case\\
& --min-anchors 5 --min-mapq 5 --bw 5000 --max-target-gap 2500 --max-query-gap 2500 & \\\bottomrule
\end{tabular}

}
\label{rs:tab:presets}
\end{table}

\subsection{Versions}\label{rs:subsec:versions}

Supplementary Table~\ref{rs:tab:versions} shows the version and the link to these corresponding versions of each tool we use in our experiments.

\begin{table}[tbh]
\centering
\caption{Versions of each tool and library.}
\begin{tabular}{@{}lll@{}}\toprule
\textbf{Tool} & \textbf{Version} & \textbf{Link to the Source Code} \\\midrule
\rs & 2.1 & \url{https://github.com/CMU-SAFARI/RawHash/releases/tag/v2.1}\\\midrule
Minimap2 & 2.24 & \url{https://github.com/lh3/minimap2/releases/tag/v2.24}\\\midrule
Dorado & 0.9.6 (for R9.4) & \url{https://github.com/nanoporetech/dorado/releases/tag/v0.9.6}\\\midrule
Dorado & 1.0.2 (for R10.4.1) & \url{https://github.com/nanoporetech/dorado/releases/tag/v1.0.2}\\\midrule
Miniasm & 0.3-r179 & \url{https://github.com/lh3/miniasm/releases/tag/v0.3}\\\midrule
Rawasm & main & \url{https://github.com/CMU-SAFARI/rawasm}\\\midrule
Flye & 2.9.5 & \url{https://github.com/mikolmogorov/Flye/releases/tag/2.9.5}\\\midrule
HERRO & 0.1 & \url{https://github.com/lbcb-sci/herro}\\\bottomrule
\end{tabular}

\label{rs:tab:versions}
\end{table}

\clearpage

\let\noopsort\undefined
\let\printfirst\undefined
\let\singleletter\undefined
\let\switchargs\undefined

\bibliographystylesupp{IEEEtran}
\setstretch{0.75}
{\small \bibliographysupp{supp}}

\end{document}